\documentclass{article}
\usepackage{spconf,amsmath,graphicx}
\usepackage{url}
\usepackage{amssymb}
\usepackage{multirow}
\usepackage{caption}
\usepackage{subcaption}
\usepackage{color}
\usepackage{tikz}
\usepackage{booktabs}
\usepackage{enumitem}
\usepackage{multirow}
\usepackage{multicol}
\usepackage{pifont}
\usepackage{hyperref}
\usepackage{tablefootnote}

\title{Emotion-Driven Melody Harmonization \\ via Melodic Variation and Functional Representation}

\name{Jingyue Huang$^1$  \quad Yi-Hsuan Yang$^2$}
\address{
         $^1$Department of Computer Science and Engineering, UC San Diego \\
	     $^2$Department of Electrical Engineering, National Taiwan University \\
         \small{\texttt{jih150@ucsd.edu, yhyangtw@ntu.edu.tw}}
	     }

\begin{document}
\ninept
\maketitle

\begin{abstract}



Emotion-driven melody harmonization aims to generate diverse harmonies for a single melody to convey desired emotions. Previous research found it hard to alter the perceived emotional valence of lead sheets only by harmonizing the same melody with different chords, which may be attributed to the constraints imposed by the melody itself and the limitation of existing music representation. In this paper, we propose a novel functional representation for symbolic music. This new method takes musical keys into account, recognizing their significant role in shaping music's emotional character through major-minor tonality. It also allows for melodic variation with respect to keys and addresses the problem of data scarcity for better emotion modeling. A Transformer is employed to harmonize key-adaptable melodies, allowing for keys determined in rule-based or model-based manner. Experimental results confirm the effectiveness of our new representation in generating key-aware harmonies, with objective and subjective evaluations affirming the potential of our approach to convey specific valence for versatile melody.
\end{abstract}

\begin{keywords}
Melody harmonization, symbolic music generation, data representation, musical keys, emotion
\end{keywords}

\vspace{-0.1cm}
\section{Introduction}
\vspace{-0.1cm}
Inspired by the remarkable achievements made in symbolic music generation \cite{remi, musictransformer, figaro, musecoco}, there is a growing interest in controlling high-level musical features during the generation process, especially the perceived emotions from music. Recent years have witnessed many efforts in unconditional music generation \cite{emopia, emogen, emomusictv, learningto, generatinglead} and melody harmonization \cite{emotion-driven, lhvae} to condition their generation on emotion. 

According to Russell’s famous Circumplex model of affect \cite{russell}, emotion could be represented in a two-dimensional space defined by valence and arousal, where \emph{valence} related to the positiveness of an emotion and \emph{arousal} refers to energy or activation \cite{mer}. Although many works are capable to control the arousal of music, few of them succeeded in controlling the perceived valence. For example, the piano music generation model in EMOPIA \cite{emopia} fails in generating low valence (i.e., negative) music, and the melody harmonization model LHVAE \cite{lhvae} found it hard to change the overall emotion of music (negative, neutral and positive) by only altering chords.

\begin{figure}
    \centering
    \includegraphics[width=\columnwidth]{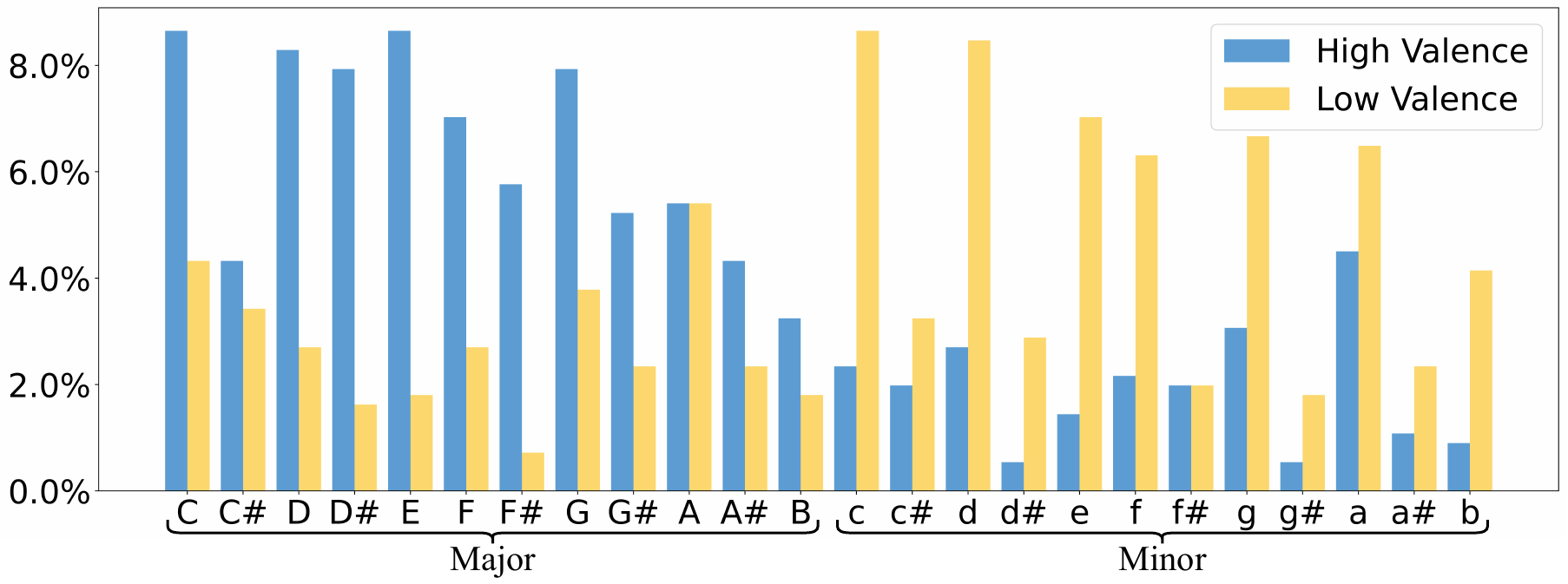}
    \caption{Key histogram of high/low valence clips from EMOPIA \cite{emopia}.}
    \label{fig:key}
\vspace{-0.5cm}
\end{figure}

The ignorance of musical keys when modeling music data may be responsible for the poor valence control. Valence is often found to be related to major-minor tonality \cite{audiofeatures}, and keys play important roles in affecting such tonality. The histogram of keys derived from the emotion-labeled music dataset EMOPIA \cite{emopia} (Fig. \ref{fig:key}) provides further support from a data perspective, where the distribution skews to major keys for high valence clips and opposite trend for low valence ones. However, to the best of our knowledge, none of existing music generation works attempted to model keys explicitly. 

In this work, we focus on the \emph{emotion-driven melody harmonization} task, which aims to convey desired emotions through harmonizing diverse chord progressions for a single melody, resulting in the creation of lead sheets. This simplified music format free from performance variances enable us to dive into the impact of keys and chords on emotions, and we only consider the valence aspect of emotion since arousal is usually related to performance-level attributes and accompaniment patterns \cite{generationof}. This task is more challenging than conventional melody harmonization from three aspects. Firstly, symbolic music datasets with both high-quality emotion and chord labels are relatively scarce. Secondly, there are even less training examples where one melody is accompanied with multiple chord progressions to convey varying emotions. Thirdly, because the key is predetermined given a melody line, there is limited room for chord progressions to contribute to the target emotion. While the first challenge was partially addressed by previous works \cite{emotion-driven, lhvae}, the later two hinder the further improvement of valence control.

Observing the above, we propose a novel functional representation designed as an alternative to REMI \cite{remi}, a popular event representation that uses note pitch values and chord names to encode symbolic music. Our method represents both melody notes and chords with \emph{Roman numerals} relative to musical keys, a \emph{functional} format considering the relationships between notes, chords and scales (major or minor) \cite{functionalharmony}. In the harmonization process, driven by an emotional condition (positive or negative), a key is determined in a rule-based or model-based manner. Subsequently, a melody line encoded in functional representation is accompanied to generate key-aware functional harmonies, a process facilitated by a Transformer model.

Compared to note pitch values and chord names, which necessitate models to infer keys implicitly, functional representation empowers us to explicitly inform musical keys while modeling music pieces. Additionally, as melodies across various scales are designed to be represented by the same set of symbols for twelve scale degrees, the likelihood of encountering two melodies with similar representations increases, and their accompanying chord progressions and emotion labels can be regarded as paired samples, offering a solution to the issue of data scarcity discussed above. Our representation also supports transposition between parallel keys\footnote{\url{https://en.wikipedia.org/wiki/Parallel_key}} for any music pieces. Specifically, when transposing to a parallel key, the scale degrees of notes and chords remain consistent, while pitches may be adjusted to reflect the new key mode, i.e., \emph{melodic variation}. Through experiments, we aim to answer the two research questions:
\begin{itemize}[leftmargin=*, itemsep=1pt, topsep=2pt]
    \item \textbf{RQ\#1}: Can the proposed representation effectively model musical keys and yield satisfactory harmonization outcomes?
    \item \textbf{RQ\#2}: Is it possible to generate different music variants from a single melody to influence the perceived valence?
\end{itemize}

Our model implementations and checkpoints are open-source\footnote{Code: \url{https://github.com/Yuer867/EMO_Harmonizer}}, and generated samples could be found in the demo webpage\footnote{Demo: \url{https://emo-harmonizer.github.io/}}.

\begin{figure}[t]
    \centering
    \includegraphics[width=\columnwidth]{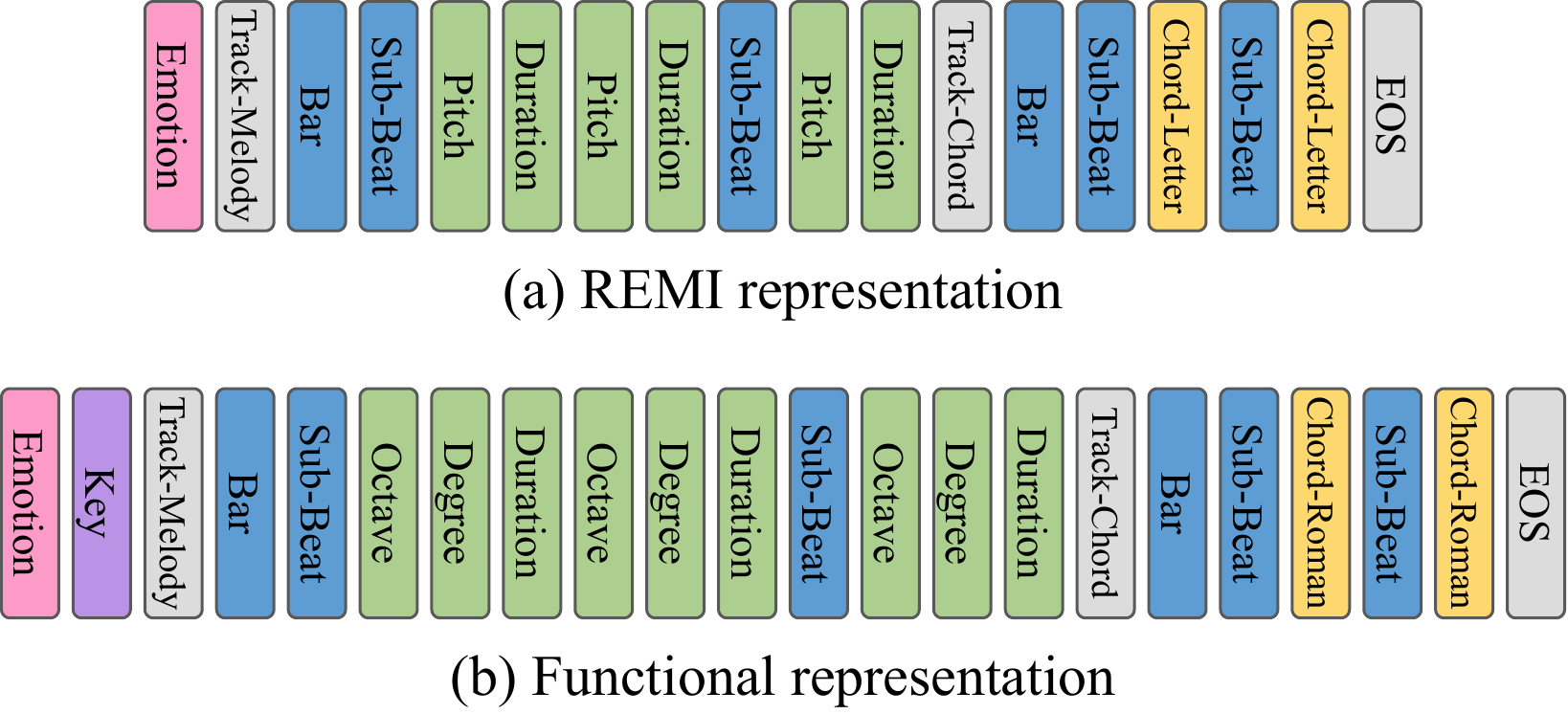}
    \caption{Illustration of (a) REMI \cite{remi} and (b) the proposed functional representation, differing in note pitch and chord name events.}
    \label{fig:representation}
\vspace{-0.5cm}
\end{figure}

\vspace{-0.2cm}
\section{Related Work}
\vspace{-0.1cm}
Various model architectures and chord control methods have been explored in melody harmonization \cite{mtharmonizer, orderlessnade, surprisenet, dat-cvae} and yielded remarkable results. In contrast, progress in emotion-driven melody harmonization is limited due to its inherent challenges. Some approaches
relied on crowd-sourced emotion tags as weak labels to obtain music samples with both emotion and chord labels \cite{emotion-driven}, while 
some approaches
manually calculated valence simply from chord labels \cite{lhvae}. For better emotion annotations, we use a pop piano performance dataset with human-annotated emotions \cite{emopia} and apply rule-based algorithm to extract chord labels. We also leverage a high-quality lead sheet dataset for pre-training. However, existing methods still struggle with controlling emotions effectively, as the discrepancy between the perceived and target emotions remain large. Moreover, none of these works consider the influence of musical keys in harmonization. They typically transpose all samples to C major\,/\,c minor for simplicity \cite{mtharmonizer,orderlessnade}, or to all 12 keys for augmentation \cite{dat-cvae}. These challenges and constraints 
motivate us to study functional representation for better emotion controllability.


\vspace{-0.2cm}
\section{Method}
\vspace{-0.1cm}
\subsection{Functional Representation}
\vspace{-0.1cm}
The proposed functional representation is designed based on REMI \cite{remi}, a widely used event (token) based representation for symbolic music, but with different note and chord events assisting to model the emotion and key information better. See Fig. \ref{fig:representation} for illustrations.

\textbf{Emotion Events} To denote distinct valences and affect overall properties, we begin the event sequence with \texttt{Emotion\_Positive} or \texttt{Emotion\_Negative} to indicate the emotional character of music clips, following the approach introduced in CTRL \cite{CTRL}.

\textbf{Key Events} After the \texttt{Emotion\_*} event, a \texttt{KEY\_*} event is appended to indicate the key property. A total of 24 types (12 tonic notes with two modes each) are used in this work. 

\textbf{Bar, Sub-Beat and EOS Events} The same as REMI, a \texttt{BAR} event is used when a new bar begins, a \texttt{SUB-BEAT\_*} event points to one of 16 possible discrete locations in a bar, and an \texttt{EOS} event will end the whole lead sheet.

\textbf{Chord Events} In existing melody harmonization studies, chord names are commonly employed to denote root notes and chord qualities. For example, \texttt{Fmaj} represents the chord \texttt{F-A-C} with root \texttt{F} and major quality. These symbols are useful when all music pieces share the same tonic, i.e., C major and c minor, but they overlook the variations in chord functions of the same chord across different keys due to different scale degrees. For example, while \texttt{Fmaj} serves tonic function in F major, it takes on subdominant function in C major. Moreover, the chord progressions following specific functional harmony rules establish tonality and convey musical emotion \cite{functionalharmony}.

\begin{figure}
    \centering
    \includegraphics[width=\columnwidth]{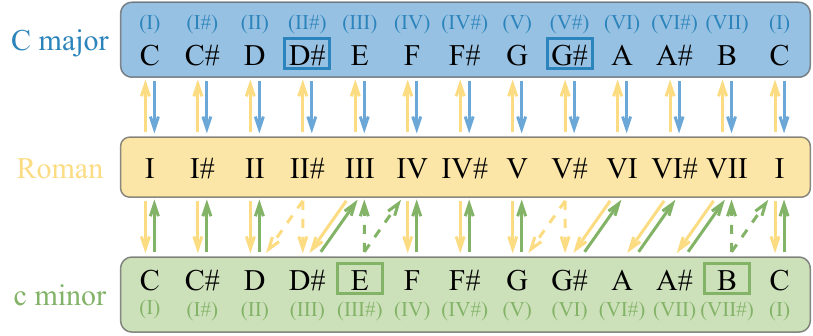}
    \caption{Illustration of the conversion between letters and Roman numerals in the cases of C major\,/\,c minor. The solid line represents a direct one-to-one conversion, while the dotted line stands for a random conversion to either one of them.}
    \label{fig:switch}
\vspace{-0.5cm}
\end{figure}

To introduce chord functions for better harmonization, we adopt Roman numerals adopted from Roman Numeral Analysis \cite{notallroads} to notate chord roots independent of keys (see Fig. \ref{fig:switch}). Given the key event, root notes belonging to the key scale (such as \texttt{C}, \texttt{D}, \texttt{E}, \texttt{F}, \texttt{G}, \texttt{A} and \texttt{B} in C major) are directly converted into Roman numerals based on their scale degrees relative to the tonic. For roots outside the scale, we employ a direct conversion for \texttt{I\#}, \texttt{II\#}, \texttt{IV\#}, \texttt{V\#} and \texttt{VI\#} appearing in major keys, but randomly assign \texttt{III\#} and \texttt{VII\#} who only appear in minor keys as one of their neighboring degrees. The conversion from Roman numerals back to letters follows a similar process. This design ensures the notation remaining key-independent and allows for harmonious transposition to any desired key while minimizing pitch variations resulting from the conversion. The notations of chord qualities remain unchanged. A \texttt{CHORD\_*} event appears every beat, even when there are no chord changes.

\textbf{Note-related Events} In REMI, every single note is denoted by 
\texttt{Pitch\_*} and \texttt{Duration\_*} events. The former indicates the onset of pitches, ranging from 21(A0) to 108(C8). To enable key-aware pitches and melodic variation, we decompose \texttt{Pitch\_*} event into \texttt{Octave\_*} and \texttt{Degree\_*} events, where degrees are computed as chord roots and octaves match the tonic's octave in the key scale. For example, pitch 63(D\#4) will be decomposed into \texttt{Octave\_4} and \texttt{Degree\_III} in C minor. Note that this work omits performance-related events such as \texttt{VELOCITY\_*} and \texttt{TEMPO\_*}.

\begin{figure}
    \centering
    \includegraphics[width=\columnwidth]{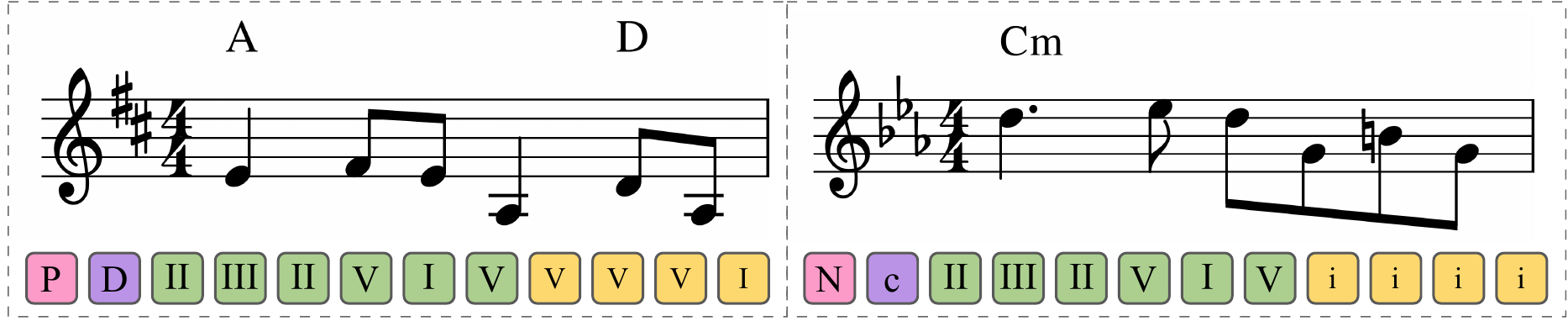}
    \caption{Examples of functional representations for two bars, featuring solely emotion, key, note pitches, and chords for simplification.}
    \label{fig:samples}
\vspace{-0.5cm}
\end{figure}

Fig. \ref{fig:samples} shows the functional representations of two bars taken from different songs in EMOPIA. Two melody lines are represented in the same formats under their corresponding keys (`\texttt{D}' major and `\texttt{c}' minor), but harmonized with different chords to convey distinct emotions (`\texttt{P}'ositive and `\texttt{N}'egative). The one on the right hand side can be considered as a melodic variation of the left one.

\vspace{-0.2cm}
\subsection{Model Architecture} \label{sec:model}
\vspace{-0.1cm}
We follow the conditional generation framework proposed in Compound Word Transformer \cite{cp} for our task. We firstly predict a key event $k$ conditioned on an emotion event $e$, and then generate a chord sequence $C$ by accompanying a melody sequence $M$ with $e$ and $k$, which could be formalized as $p(k, C|e, M) = p(k|e)p(C|e, k, M)$.
A sequence-to-sequence model \cite{cp, c&e} is applied to learn $p(k|e)$ and $p(C|e, k, M)$ simultaneously. Moreover, with the positions of \texttt{BAR} events, $M$ and $C$ are further segmented into $\{M_1, \cdots, M_b\}$ and $\{C_1, \cdots, C_b\}$, where $b$ is the number of bars. The segmented sequences are interleaved in the form of $\{ \cdots $\texttt{TRACK\_MELODY}, $M_i$, \texttt{TRACK\_CHORD}, $C_i \cdots \}$ with additional \texttt{TRACK\_*} events, so that the target chord bar $C_i$ is nearest to its corresponding conditions $M_i$ and the dependency between melody and chord is easier to learn. The final model is summarized as
$p(k, C|e, M) = p(k|e) \prod_{i=1}^b p(C_i|e, k, M_{\leq i}, C_{<i})$
and minimized the negative log-likelihood loss of the generated sequence.

At inference time, given an emotion condition $e$, a key event $k$ could be determined in rule-based or model-based manner. In the former approach, we enforce the original key to its parallel major key for positive emotion or parallel minor key otherwise, inspired by Fig. \ref{fig:key}. However, a song in major key could also convey negative emotion, so alternatively, the key can be predicted from $p(k|e)$ learned above. Note that the corresponding melody will be adjusted as key changes. 

As the size of the dataset with emotion labels is not big enough, we pretrain the model with a large lead sheet dataset without emotion annotations to establish a robust understanding of relationship between melody and chord, where \texttt{EMOTION\_NONE} is used as the emotion event. We then finetune the model on EMOPIA \cite{emopia}
to learn harmonization styles specific to different emotion conditions.

\vspace{-0.2cm}
\section{Experiments}
\vspace{-0.1cm}
\subsection{Datasets, Preprocessing, and Model Settings}
\vspace{-0.1cm}
We adopt a large 
lead sheet dataset collected from HookTheory \cite{hooktheory}
released in SheetSage \cite{sheetsage} for pre-training (refered to as ``HookTheory" hereafter). Each piece in HookTheory contains high-quality, human-transcribed melody, chord and key annotations. After removing the pieces that are not in 4/4 time signature or major\,/\,minor\,keys, over 18k segments remain. We simplify 249 chord quality classes to 11 types including major, minor, augment, diminish, suspend2, suspend4, major7, minor7, dominant7, diminish7, half-diminish7, 
which may be sufficient to convey most emotional characters.

We use the 
piano MIDI dataset
EMOPIA for model finetuning. It contains 1,071 music clips 
with human-annotated emotion labels. A four-class taxonomy adopted from Russell's model \cite{russell} is used for annotations, including HVHA (high valence high arousal), HVLA (high valence low arousal), LVHA (low valence high arousal) and LVLA (low valence low arousal). Since we only consider the valence dimension, clips in HVHA and HVLA are combined into Positive class and others into Negative class. To form the lead sheet samples from piano performance, we adopt a method similar to the one proposed in \cite{chordalanalysis} to extract chord labels in 11 quality types, and apply a heuristic rule-based 
method to extract the 
melody line \cite{miditoolkit}.

The statistics of two datasets after preprocessing are shown in Table \ref{table:dataset}. The clips in HookThoery are randomly divided into train and validation splits, and we follow the stratified split code provided by EMOPIA for train and validation, both with the ratio of 9:1. In our data representations, the vocabulary size of events is 217.

\begin{table}[t]
\centering
\resizebox{0.8\columnwidth}{!}{
\begin{tabular}{lrcc}
\toprule
\textbf{Dataset} & \textbf{\# clips} & \textbf{\# bars (avg.)} & \textbf{\# events (avg.)} \\
\midrule
HookTheory\,\cite{sheetsage}       & 18,206            & 10.84           & 339.27            \\
EMOPIA\,\cite{emopia}          & 1,071             & 16.96           & 591.80            \\
\bottomrule
\end{tabular}
}
\caption{Summary of datasets used in our experiments.}
\label{table:dataset}
\vspace{-0.6cm}
\end{table}

A 12-layer linear Transformer with Performer attention \cite{performer} is used for generation (8 attention head, 512 hidden state dim., 38 million parameters) and trained with a batch size of 4 and a maximum sequence length of 1,024 
(longer than 99.2\% and 94.6\% of clips in HookTheory and EMOPIA respectively) 
on a Tesla V100 GPU with 32G VRAM. We use Adam optimizer with 200 steps of warmup to maximum learning rate of 1e--4 and 1e--5 for pretrain and finetune respectively, both followed by 500k steps of cosine decay. Checkpoints with lowest validation loss are  used for inference, using nucleus sampling \cite{sampling} with temperatured softmax ($\tau = 1.1$, $p = 0.99$). 

\vspace{-0.2cm}
\subsection{Objective Evaluation}
\vspace{-0.1cm}

We firstly study \textbf{RQ\#1}, i.e., can the proposed representation effectively model musical keys and yield satisfactory harmonization outcomes. We consider two baselines for comparison. One represents music clips in REMI \cite{remi} directly without any other processing, following the approach used in emotion-conditioned melody harmonization work LHVAE \cite{lhvae}. The other transposes (`\textit{trans}') all clips to C major\,/\,c minor before REMI encoding, standing for a common method of existing melody harmonization works \cite{mtharmonizer,orderlessnade}. We do not compare with other model architectures as our primary focus is to study the impact of different representations. For functional representation, besides the full format, we also compare one ablated version (`\textit{ablated}'), where chords are encoded in the functional format but melody notes are represented by note pitch values as in REMI.

For each of the variants, we harmonize all melodies in validation set under the conditions of positive and negative emotion respectively, i.e., $88 \times 2$ samples each variant, to check their harmonization performance. Two groups of objective evaluation metrics are considered, including three metrics proposed in \cite{mtharmonizer} to evaluate the harmonicity between chord and melody: chord tone to non-chord tone ratio (CTnCTR), pitch consonance score (PCS), melody-chord tonal distance (MCTD), and two metrics \emph{newly} proposed here for evaluating the ability of modeling keys:

\begin{itemize}[itemsep=0pt, topsep=0pt]
    \item \textbf{Root ratio} (RR): the ratio of chord \emph{roots} in key scale.
    \item \textbf{Note ratio} (NR): the ratio of chord \emph{notes} in key scale.
\end{itemize}

Table \ref{table:harmonization} shows that, 
while encoding music clips into REMI representation directly have poor performances in all metrics, providing musical keys information by transposing to C major\,/\,c minor or adding key events with functional representation could greatly improve their harmonicity and almost meet the real data. This  indicates that musical key is critical to melody harmonization,
and that our representations enables efficient learning of their relationships despite of fewer training data for each key type. Moreover, the ablated version suffer losses on melody-related metrics, showing the necessity of encoding melodies in functional formats. In short, the answer to RQ\#1 is that when applying functional representation on both melody notes and chord labels, musical keys could be effectively modeled and the melodies are well harmonized.

\begin{table}[t]
\centering
\resizebox{0.85\columnwidth}{!}{
\begin{tabular}{l|ccc|cc}
\toprule
\textbf{Methods}    & \textbf{CTnCTR}     & \textbf{PCS}        & \textbf{MCTD}         & \textbf{RR}              & \textbf{NR}           \\ \midrule
REMI\,\cite{remi,lhvae}               & 0.291               & 0.211               & 1.731                 & .647                    & .592                 \\
REMI (\textit{trans})      & 0.747               & \textbf{1.557}      & 1.347                 & \textbf{.927}                    & \textbf{.923}        \\
\midrule
Ours                & \textbf{0.750}      & 1.487               & \textbf{1.343}        & \textbf{.943}           & .935               \\
Ours (\textit{ablated})     & 0.530               & 0.880               & 1.501                 & .952                    & .942                 \\
\midrule
Real data           & 0.801               & 1.613               & 1.314                 & .935                    & .926                 \\
\bottomrule
\end{tabular}
}
\caption{Melody harmonization objective evaluation results (the closer to the real data, the better).}
\label{table:harmonization}
\end{table}

\begin{table}[t]
\centering
\resizebox{\columnwidth}{!}{
\begin{tabular}{l|cc|ccc}
\toprule
\multirow{2}{*}{\textbf{Methods}}  & \multicolumn{2}{|c|}{\textbf{Objective $\downarrow$}}      & \multicolumn{3}{|c}{\textbf{Subjective $\uparrow$}} \\
                                   & \textbf{QD} & \textbf{PD} & \textbf{Novice} & \textbf{Expert} &  \textbf{Total}       \\
\midrule
REMI (\textit{trans})                    & .0128           & .0218      & $-$0.04\footnotesize{$\pm$0.92}  &  $-$0.33\footnotesize{$\pm$0.94}           & $-$0.16\footnotesize{$\pm$0.94} \\
REMI (\textit{rule})                  & .0140           & .0161      & ~~0.44\footnotesize{$\pm$1.12}   &  ~~\textbf{0.69\footnotesize{$\pm$1.08}}  & ~~0.54\footnotesize{$\pm$1.11}  \\
\midrule
Ours                               & .0127           & .0119      & ---                 & ---                 & ---                \\
Ours (\textit{rule})                 & \textbf{.0084}  & \textbf{.0116} & ~~~~\textbf{0.61\footnotesize{$\pm$0.91}} & ~~0.58\footnotesize{$\pm$1.11} & \textbf{0.60\footnotesize{$\pm$1.00}}              \\
Ours (\textit{model})                  & .0114           & .0130      & $-$0.093\footnotesize{$\pm$1.36} & ~~0.39\footnotesize{$\pm$1.16}  & 0.10\footnotesize{$\pm$1.31}  \\
\bottomrule
\end{tabular}
}
\caption{Emotion controllability results via objective evaluation (the lower the better) and subjective evaluation (the higher the better).}
\label{table:controllability}
\vspace{-0.5cm}
\end{table}


Next, we study \textbf{RQ\#2}, i.e., whether we could control the valence when generating different music variants from a  melody.
We select the methods perform well above, i.e., REMI\,(\textit{trans}) and our functional representation.
Besides simply setting the emotion event to the target emotion without changing keys, we further examine rule-based (`\textit{rule}') and model-based (`\textit{model}') methods discussed in Section \ref{sec:model} to determine keys conditioned on emotions, and build five models in total. Note that the `\textit{rule}' variant of REMI\,(\textit{trans}) is implemented by transposing the melody to major\,/\,minor keys directly. To quantify how well the generated samples conform to the emotion conditions, we propose the following two metrics:
\begin{itemize}[leftmargin=*, itemsep=0pt, topsep=1pt]
    \item \textbf{Quality Distribution} (QD): compute the KL divergence between the chord quality distributions of generated samples and real data for positive and negative emotions respectively, then take average.
    \item \textbf{Progression Distribution} (PD): calculate the KL divergence between the bi-gram chord progression distributions as above and compute the average. Progression refers to the difference between two consecutive chord root in chromatic scale. For example, the progression between \texttt{Dmaj} and \texttt{Fmin} is 3.
\end{itemize}

To obtain reliable distributions, for each combination, we generate harmonies for all melodies in validation set five times under both positive and negative emotion conditions, i.e., $88 \times 5 \times 2$ samples, to match the size of training samples. As is shown in the left part of Table \ref{table:controllability}, the combination of functional representation and rule-based key determination achieves the lowest distribution distances for two metrics. Furthermore, all the versions using functional representation performs better than REMI ones, indicating the effectiveness of this representation to model emotion-related properties of music. 

\vspace{-0.2cm}
\subsection{User Study}
\vspace{-0.1cm}
An online survey was deployed to collect user responses. Every subject needs to listen to 16 music pieces, including two original pieces in positive and two in negative as well as their four variants generated with the same melody but the opposite emotion condition by four methods\footnote{The method using functional representation without key changes is removed here to lower the burden of users} introduced above. 
These original pieces are randomly drawn from the validation set. For each group of samples, users will 
assess how the variants (in random order) compare to the original piece in terms of their emotional differences on a five-point scale, specifically whether they convey a much more positive(2), more positive(1), unchanged(0), more negative(--1) or much more negative(--2) emotion, without knowing the pre-defined emotion conditions. This design is inspired by the finding that it is easier for human subjects to make relative valence comparison rather than to assign absolute rating \cite{ranking-based}. 23 subjects participated in the survey (`Total'), 9 of them with $\geq$4 years of musical training or experience (`Expert'). 

The right side of Table \ref{table:controllability} shows the average scores from the subjects, with
a higher score ($\in [-2,2]$) indicating better emotion controllability, i.e., the perceived valence matches the given condition. Totally speaking, the combination of functional representation and rule-based key determination performs best, followed by REMI with rule-based keys, which indicates that musical keys play a significant role in influencing the perceived valence through melodic variation and the functional representation generates slightly better harmonies to support the perception of desire emotions, which answers the RQ\#2. Moreover, users are still unable to perceive the emotion conversions by simply harmonizing the original melody without any key changes, aligning with the previous findings \cite{lhvae}. When determining keys in model-based method, sometimes major (minor) keys will be sampled for negative (positive) emotions and the generated harmonies lack the ability to obviously change the perceived valence by themselves, yielding worse emotion controllability. Namely, the rule-based way seems to outperform the model-based way in determining keys. Improving the latter will be a focus of future work.

\begin{figure}
    \centering
    \includegraphics[width=0.85\columnwidth]{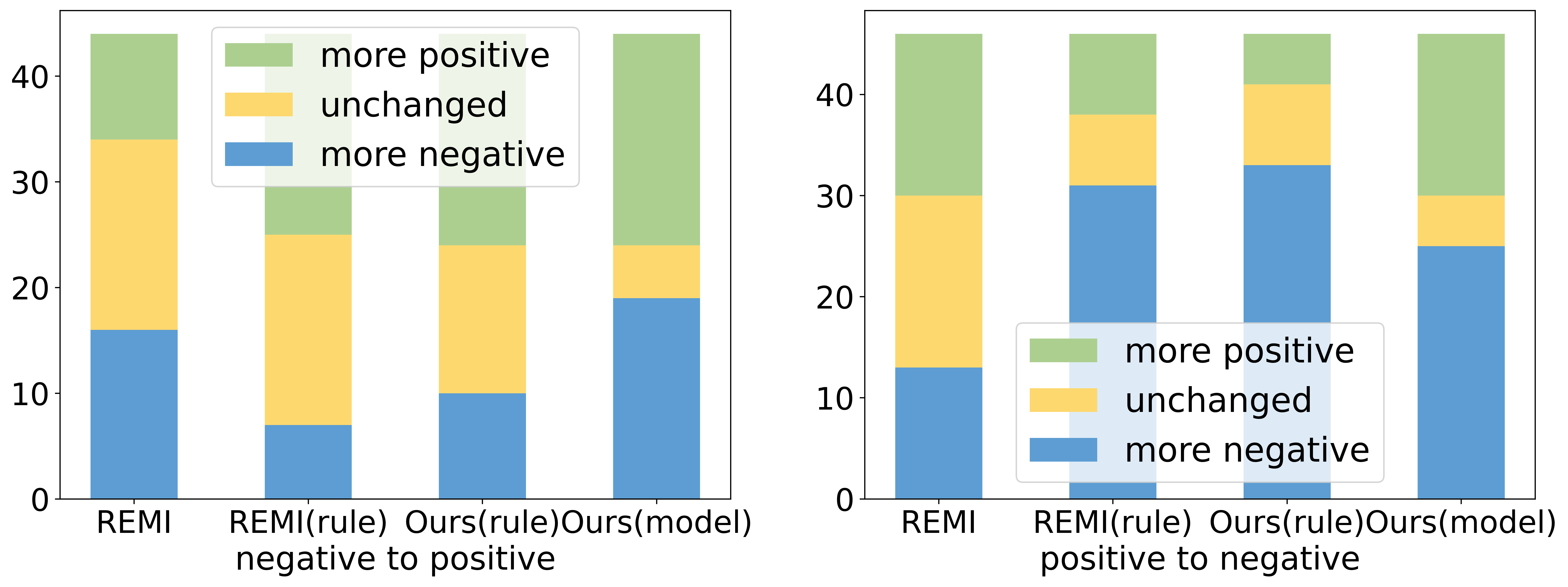}
    \vspace{-0.2cm}
    \caption{Comparison of subjective results under different settings.}
    \label{fig:emotion}
\vspace{-0.6cm}
\end{figure}

When diving into user responses from different musical backgrounds, those with longer years of music-related experiences seem to favor the REMI-generated samples, while people with less experiences choose the functional representation. Intuitively, with more musical training, people may analyze the conveyed emotions more from the perspective of music theory, while novices simply rely on their emotional senses, but it is hard to say which method is more accurate. Moreover, if examining the cases of `negative to positive' (i.e., generating positive variants for negative original pieces) and `positive to negative' separately (Fig. \ref{fig:emotion}), the control of low valence is better with more samples match the negative condition, while half of samples conditioned on positive are still identified as negative ones as their ground-truth emotions. It seems that minor keys have significant influence on negative feelings, while the melody flows have strong influences on positive moods.

\vspace{-0.2cm}
\section{Conclusion and Future Work}
\vspace{-0.2cm}
In this paper, we have proposed a novel functional representation for symbolic music, which represents melody notes and chords with Roman numerals relative to musical keys. A Transformer-based framework is then adopted to harmonize melodies conditioned on emotional valence. Objective assessments validate our approach's effectiveness in key modeling, while subjective evaluations confirm its ability to convey desired emotional valence. Future endeavors may focus on the improvement of model-based approach and the control of arousal aspect of emotion through accompaniment generation.

\bibliographystyle{IEEEbib}
\bibliography{refs}

\end{document}